\documentclass[showpacs,preprintnumbers,amsmath,amssymb,twocolumn,superscriptaddress,prl]{revtex4}
\usepackage{graphicx}
\usepackage{amsfonts}
\usepackage{amsmath, amsthm, amssymb}
\usepackage{dsfont}
\usepackage{mathptm}

\newcommand{\ve}[1]{\mathbf{#1}}

\newcommand{\vk}{\ve{k}} 
\newcommand{\vp}{\ve{p}} 

\newcommand{\vpf}{\mathbf{\vp}_\text{F}} 

\newcommand{\vg}{\ve{g}} 

\newcommand{\eg}{\textit{e.g. }}
\newcommand{\etal}{\emph{et al.}}
\def\i{\mathrm{i}}

\begin{document}
\title[Spin-sensitive Long-ranged Proximity Effect for Triplet Superconductors]{Spin-sensitive Long-ranged Proximity Effect for Triplet Superconductors}

\author{Gaetano Annunziata}
\affiliation{SPIN-CNR, I-84084 Fisciano (Salerno), Italy \\
Dipartimento di Fisica ``E. R. Caianiello'', Universit\`a di
Salerno, I-84084 Fisciano (Salerno), Italy}

\author{Mario Cuoco}
\affiliation{SPIN-CNR, I-84084 Fisciano (Salerno), Italy \\
Dipartimento di Fisica ``E. R. Caianiello'', Universit\`a di
Salerno, I-84084 Fisciano (Salerno), Italy}

\author{Canio Noce}
\affiliation{SPIN-CNR, I-84084 Fisciano (Salerno), Italy \\
Dipartimento di Fisica ``E. R. Caianiello'', Universit\`a di
Salerno, I-84084 Fisciano (Salerno), Italy}

\author{Asle Sudb{\o}}
\affiliation{Department of Physics, Norwegian University of Science and Technology, N-7491 Trondheim, Norway}

\author{Jacob Linder}
\affiliation{Department of Physics, Norwegian University of Science and Technology, N-7491 Trondheim, Norway}

\date{Received \today}
\begin{abstract}
The discovery of noncentrosymmetric superconductors, such as CePt$_3$Si, and chiral superconductors, such as Sr$_2$RuO$_4$, calls for experimental methods to identify the presence of spin-triplet pairing. We here demonstrate a method which accomplishes this in an appealingly simple manner: a spin-sensitive proximity effect in a ferromagnet$\mid$triplet superconductor bilayer. It is shown how the orientation of the field can be used to unambiguously distinguish between different spin-triplet states. Moreover, the proximity effect becomes long-ranged in spite of the presence of an exchange field and even without any magnetic inhomogeneities, in contrast to conventional S$\mid$F junctions. Our results can be verified by STM-spectroscopy and could be useful as a tool to characterize the pairing state in unconventional superconducting materials.

\end{abstract}

\maketitle

Superconductivity may be defined as conventional
or unconventional depending on the properties of the pairing state and whether or not multiple broken
symmetries are present in the system. In conventional superconductors, the pairing state belongs to the
trivial representation of the point-group and the system ground state breaks the
U(1) gauge symmetry. On the other hand, unconventional superconductors display pairing
symmetries belonging to higher-dimensional representations of the point group and
may also exhibit multiple broken symmetries in the ground state. Examples of the latter
are ferromagnetic superconductors \cite{saxena_nature_00} with simultaneously broken U(1) and SU(2) symmetries
due to the presence of an intrinsic magnetization, and noncentrosymmetric superconductors \cite{bauer_prl_04},
where the absence of a definite parity of the lattice leads to the mixing of even- and odd-parity
superconducting order parameters.

\par
In order to characterize the properties of a superconducting system, much relies on
determining the orbital- and spin-symmetry of the order parameter (OP), a topic currently
under intense investigation \cite{nelson_science_04,yuan_prl_06,curro_nature_05,lebed_prl_06,mazin_prl_05}.
To acquire information about the order parameter, it is often useful to study how the superconducting correlations behave when placed
in proximity to a non-superconducting material such as a normal metal. This idea has been employed previously in several works studying \textit{e.g.} normal metal$\mid$non-centrosymmetric superconductor (N$\mid$NCS) junctions \cite{yokoyama_prb_05, linder_prb_07, iniotakis_prb_07, vorontsov_prl_08} in order to look for unique signatures of the superconducting OP. However, the non-superconducting material does not necessarily have to be a simple normal metal. Instead, it may feature intrinsic properties, such as magnetism, which then provide an arena for studying the interplay between superconductivity and different types of electronic ordering \cite{interplay}.
\par

\begin{figure}[th!] \centering \resizebox{0.35\textwidth}{!}{
\includegraphics{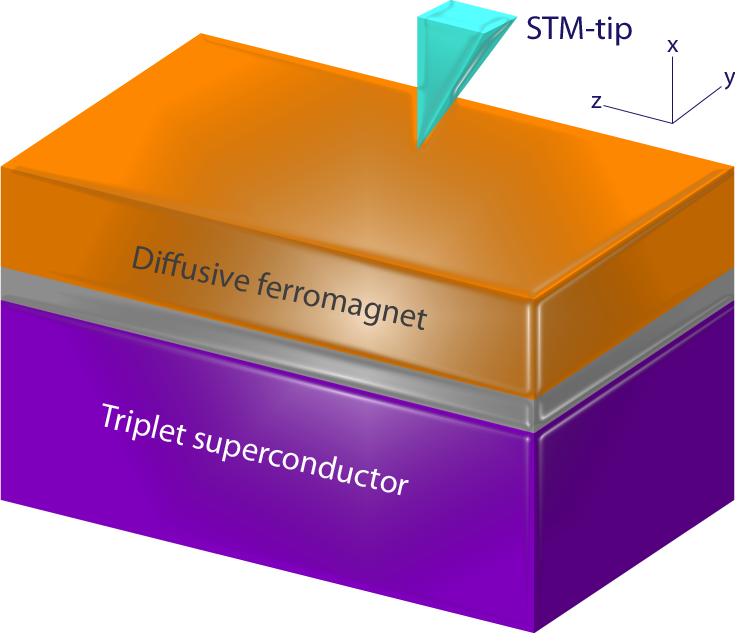}}
\caption{(Color online). The experimental setup proposed for a
spin-sensitive proximity
 effect: a ferromagnet$\mid$triplet superconductor bilayer separated by a thin insulating barrier. The ferromagnetic layer has a thickness $d_F$ and the superconducting condensate is characterized by a $\mathbf{d}_\vk$-vector.}
 \label{fig:model}
\end{figure}


A natural question arises: could such an interplay be useful in order to extract information about the superconducting state? In this Letter, we demonstrate that a ferromagnet$\mid$triplet superconductor bilayer provides an appealingly simple and powerful method of clearly distinguishing between different types of triplet pairing states, thus providing information about the nature of the superconducting condensate. We show that this is a direct result of a \textit{spin-sensitive long-ranged proximity effect}. The interesting features about this effect are that it provides \textit{i)} unambiguous signatures in the DOS due to an interplay between the exchange field $\mathbf{h}$ and the $\mathbf{d}_\vk$ vector and also that \textit{ii)} the proximity effect becomes long-ranged in spite of the presence of $\mathbf{h}$, thus decaying on a scale of the normal coherence length $\xi_T = \sqrt{D/T}$ where $D$ is the diffusion constant and $T$ is the temperature. In this way, one may in a controllable way probe experimentally the nature of the pairing symmetry in unconventional superconducting materials via \textit{e.g.} scanning-tunneling-microscopy (STM) measurements in the proximate non-superconducting region. Moreover, this finding suggests an alternative to other methods where measurements are performed on the superconductor itself in the presence of an external magnetic field, which can lead to ambiguous interpretations due to \eg formation of a vortex lattice. Finally, the effect predicted in this Letter constitutes a way of generating a long-range proximity effect \textit{without any magnetic inhomogeneity in the ferromagnetic layer}. This is in contrast with the situation considered in conventional S$\mid$F structures, where such a long-ranged proximity effect only occurs in the presence of inhomogeneous magnetization \cite{bergeret_rmp_05}. We explain the microscopic mechanism responsible for the discrimination between different triplet states,
 due to the coupling to the ferromagnetic exchange field $\mathbf{h}$, which is a general result that can be applied to identify the $\mathbf{d}_\vk$-vector in arbitrary triplet superconductors. As a concrete application of our results, we consider two specific cases: a non-centrosymmetric superconductor and a chiral superconductor. For these two cases, we demonstrate the manifestation of a spin-sensitive long-ranged proximity effect in the density of states.

Let us introduce the theoretical framework employed here to arrive
at our main results. We will use units such that $\hbar=c=1$ and
use $\underline{\ldots}$ for $2\times2$ matrices while
$\hat{\ldots}$ denotes $4\times4$ matrices. We will make use of
the quasiclassical theory of superconductivity, where the main
assumption is that the Fermi energy is the largest energy-scale in
the system, and sketch the framework for the specific case of a
non-centrosymmetric superconductor. So far in the literature,
there exists no study of the proximity effect for a
non-centrosymmetric superconductor. Here, we shall study precisely this and
additionally demonstrate how this proximity effect becomes
spin-sensitive when the proximity layer is a ferromagnet. The
antisymmetric spin-orbit coupling in the NCS is modelled by a
Rashba-type interaction $\mathbf{g}_\vk =
\lambda(\hat{\mathbf{n}}\times\vk),$ where $\lambda$ denotes the
strength of the spin-orbit interaction and $\hat{\mathbf{n}}$
denotes the axis of broken inversion symmetry. More specifically,
the crystallographic structure of the material does not have a
mirror plane with $\hat{\mathbf{n}}$ as normal vector. In what
follows, we will consider a situation where the mirror plane with
the [001] direction as normal vector is lost in the crystal, i.e.
$\hat{\mathbf{n}}=\hat{\mathbf{z}}$. The spin-orbit coupling
vector may then be written as $\vg_\vk = \lambda(-k_y,k_x,0).$ In
general, one then finds that the superconducting order parameter
in the original spin basis may be written as
$\underline{\Delta}_\vk = \Delta_{0} \i\underline{\sigma_y} +
\mathbf{d}_\vk\cdot\boldsymbol{\underline{\sigma}}\i\underline{\sigma_y},\;
\mathbf{d}_\vk = \Delta_{t} \left(-k_y,k_x,0 \right)/k_F,$ where
$k_F$ is the Fermi wavevector and $\Delta_{0}$ and $\Delta_t$ are
singlet and triplet gap amplitudes for NCS, respectively. The
singlet amplitude $\Delta_{0}$ will be used as reference energy.

We now obtain a general expression for the retarded Green's
function in a bulk noncentrosymmetric superconductor. Starting out
from the second quantized Hamiltonian in real-space, one finds the
following expression for the matrix Green's function in Spin
$\otimes$ Nambu space:
\begin{align}
\hat{G}(\vp,\varepsilon) &= (\varepsilon \hat{\rho}_3 -
\hat{\xi}_\vp - \hat{\Sigma}_\vp + \hat{\Delta}_\vp)^{-1},\; \hat{\xi}_\vp = [\vp^2/(2m)]\hat{1}.\notag\\
\hat{\Sigma}_\vp &= \begin{pmatrix}
\vg_\vp\cdot \underline{\boldsymbol{\sigma}} & 0\\
0 & [\vg_{-\vp}\cdot \underline{\boldsymbol{\sigma}}]^\mathcal{T} \\
\end{pmatrix}, \;\hat{\Delta}_\vp = \begin{pmatrix}
0 & \underline{\Delta}_\vp \\
\underline{\Delta}_{-\vp}^* & 0 \\
\end{pmatrix},\;
\end{align}
The quasiclassical Green's function $\hat{g}(\vp_F,\varepsilon)$
is then obtained by integrating out the dependence on kinetic
energy. The Green's function is assumed to be strongly peaked
around Fermi level, and one obtains $\hat{g}(\vp_F,\varepsilon) =
\frac{\i}{\pi} \int^\infty_{-\infty} \text{d}\xi_\vp
\hat{G}(\vp,\varepsilon).$  In the NCS, a high impurity
concentration would suppress completely the odd-parity
spin-triplet component $\mathbf{d}_\vk$ of the superconducting
order parameter. We therefore consider a ballistic superconducting
region, and make use of the bulk solution $\hat{g}_S$ which may be
obtained from the Eilenberger equation by setting the gradient
term to zero \cite{hayashi_prb_06}: $[\varepsilon\hat{\rho}_3 -
\hat{\Sigma}_{\vpf} + \hat{\Delta}_{\vpf}, \hat{g}_S]_- = 0.$ The
Green's function reads \cite{hayashi}:
\begin{align}\label{eq:bulkncs}
\hat{g}_S(\varphi) &= \frac{1}{2}\begin{pmatrix}
\underline{\mathcal{C}}(\varphi) & \underline{\mathcal{S}}(\varphi) \\
\underline{\mathcal{S}}(-\varphi) & -\underline{\mathcal{C}}(-\varphi) \\
\end{pmatrix},\; [\hat{g}_S(\varphi)]^2 = \hat{1},
\end{align}
where $\mathcal{C}(\varphi) = c^+\underline{1} +
(\underline{\sigma_y}\cos\varphi -
\underline{\sigma_x}\sin\varphi)c^-$, $\mathcal{S}(\varphi) =
\i\underline{\sigma_y}s^+ + (\underline{1}\cos\varphi +
\underline{\sigma_z}\sin\varphi)s^-$, and $\underline{\ldots}$
denotes a 2$\times$2 matrix in spin-space. Above, we defined
$c^\pm=c_+ \pm \ c_-$, $s^\pm=s_+ \pm \ s_-$, being
$c_\pm=\cosh(\theta_\pm)$, $s_\pm=\sinh(\theta_\pm)$, $\theta_\pm
= {\rm{arctanh}}(\Delta^\pm / \varepsilon)$ and
$\Delta^\pm=\Delta_0 \pm \ \Delta_t$, having chosen real
amplitudes for singlet and triplet components of the gap function.
Note that $\varphi$ is the azimuthal angle in the $xy$-plane. 
The inverse proximity effect, from the ferromagnet into
the superconductor, can be shown to be negligible for a
low-transparency barrier and wide superconducting region \cite{linder_prb_10}.

\par
In the non-superconducting region, we consider the diffusive
regime of transport which often is the experimentally most
relevant one. In order to calculate the Green's function
$\hat{g}$, we need to solve the Usadel equation with appropriate
boundary conditions at $x=0$ and $x=d_F$. Since we employ a
numerical solution, we have access to study the full proximity
effect regime. 
The Usadel equation \cite{usadel} in the N part reads:
\begin{align}\label{eq:usadel}
D\partial(\hat{g}\partial\hat{g}) + \i[\varepsilon\hat{\rho}_3 +
\text{diag}[\mathbf{h}\cdot\underline{\boldsymbol{\sigma}},
(\mathbf{h}\cdot\underline{\boldsymbol{\sigma}})^\mathcal{T}],
\hat{g}]=0.
\end{align}
Boundary conditions applicable to diffusive normal
metal$\mid$unconventional superconductor junctions were derived in
\cite{tanaka_prl_03}. We assume that materials are separated by an
infinitely thin insulating barrier of the form $H  \delta(x)$,
corresponding to a transmissivity $T(\varphi) = \frac{4 \
\cos^2(\varphi)}{4 \ \cos^2(\varphi) + \ Z^2},$ where $Z=2 m H /
k_F$ is a dimensionless parameter quantifying the barrier
strength. Other relevant parameters are the resistance of
non-superconducting layer $R_F$ and insulating barrier resistance
$R_B = 2R_0/t$, where $t=\int_{-\pi/2}^{\pi/2}d\varphi\
T(\varphi)\cos(\varphi)$ and $R_0$ is the contact Sharvin
resistance. Being it a function of constriction area, it can be
considered an independent parameter with respect to $Z$. For
arbitrary interface transparency boundary condition at $x=0$ can
be written as $\Gamma d_F  \hat{g}  \partial  \hat{g}= 2
\langle\left[\hat{g}, \hat{B}(\varphi)\right]\rangle,$ where
$\Gamma=R_B/R_F$ and $\langle\ldots\rangle$ represents an angular
average on Fermi surface: $\langle f(\varphi)\rangle =
\int_{-\pi/2}^{\pi/2}d\varphi\ f(\varphi)\cos(\varphi)/t$. The
matrix $\hat{B}(\varphi)$ can be written as
\begin{align}
\hat{B}(\varphi) = \frac{ -T'\left(\hat{1}+\hat{H}_-^{-1}
\right)+T'^2\hat{g}\hat{H}_-^{-1}\hat{H}_+}{-T'\left[\hat{g},\hat{H}_-^{-1}\right]+
\hat{H}_-^{-1}\hat{H}_+ -T'^2\hat{g}\hat{H}_-^{-1}\hat{H}_+\hat{g}
},
\end{align}
where $\hat{H}_\pm=\left(\hat{g}_S (\varphi)\pm \ \hat{g}_S
(\pi-\varphi)\right)/2$ and $T'(\varphi) =
T(\varphi)/[2-T(\varphi)+2\sqrt{1-T(\varphi)}].$ At $x=d_F$ the N
part borders to vacuum so that the boundary condition simply read
$\partial \hat{g} = \hat{0}.$

In what follows, we analyze the proximity effect in the normal
electrode by examining the quasiparticle density of
states (DOS) defined as $N(\varepsilon)/N_0 =
\text{Re}[g_{11}+g_{22}]/2$, where $N_0$ is the normal-state DOS. To be as concrete as possible, we
restrict ourselves to a realistic choice of parameter set:
$d_F/\xi_S=1.5$ (corresponding to a Thouless energy $E_\text{Th}\approx
0.45 \Delta_0$), $Z=2$, $\Gamma=0.01$, and later comment on how
their variation would affect our results. This choice ensures that
the junction has a rather low interface-transparency and that the
ferromagnetic layer has a thickness comparable to the
supercondcuting coherence length. To model inelastic scattering,
we add a small imaginary part $\delta$ to the quasiparticle
energies, where $\delta/\Delta_0 = 0.01$. The system under
consideration consists of a typical STM measurement setup where
the layers lie in the $yz$-plane, thus stacked along the $x$-axis
(see Fig. \ref{fig:model}). Our calculations assume translational
invariance in the directions parallel to the interface region and
are performed in 2D. However, we underline that the results
reported here remain equally valid in 3D due to the symmetry of
the superconducting order parameters considered.


\begin{figure}[t!] \centering \resizebox{0.45\textwidth}{!}{
\includegraphics{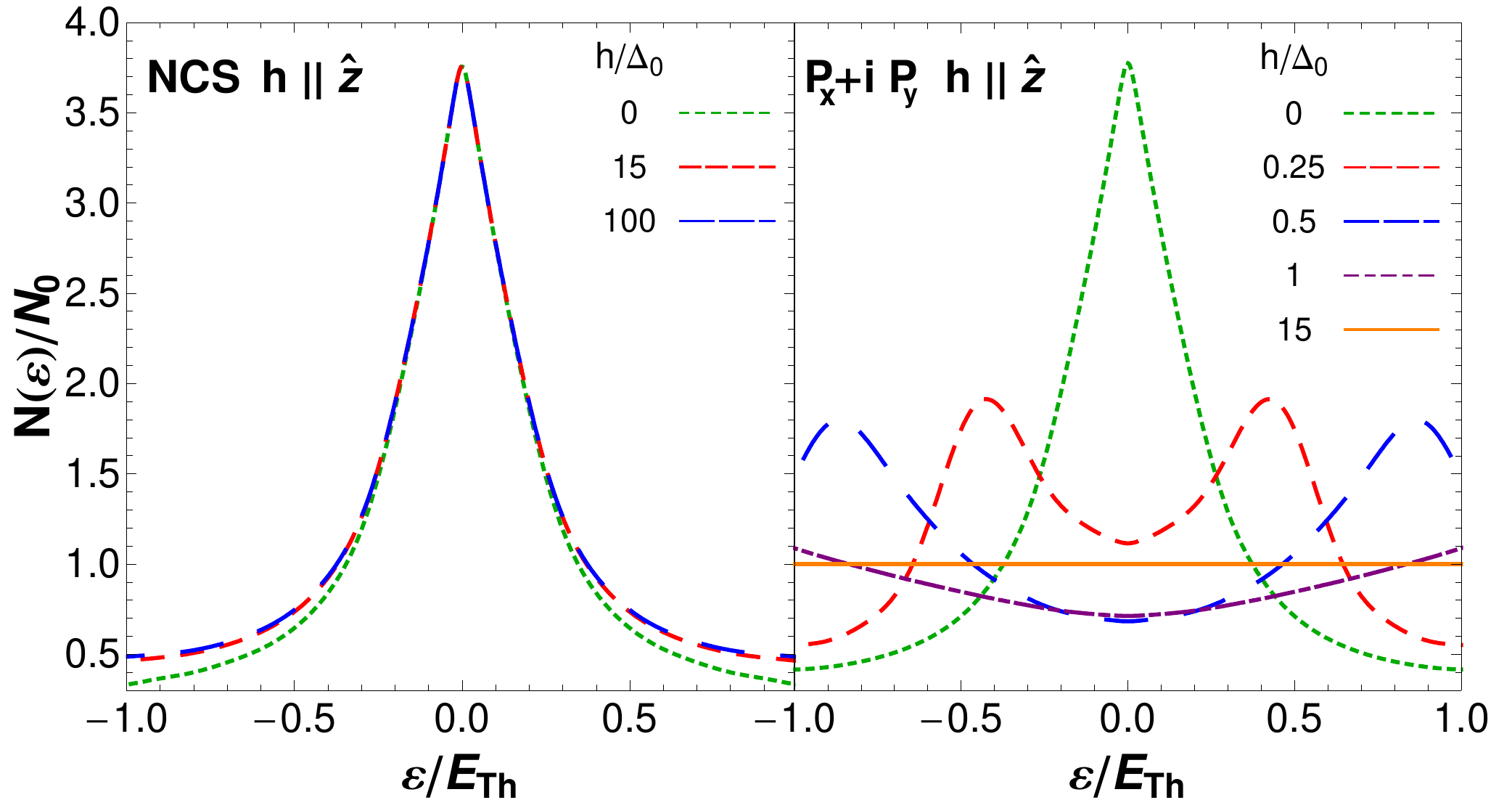}}
\caption{(Color online). {\it Left panel}: DOS at $x=d_F$ in a
F$\mid$NCS junction for ${\bf h}|| \hat{{\bf z}}$ and different
values of exchange field. {\it Right panel}: DOS at $x=d_F$ for
chiral $p_x+ip_y$ superconducting electrode for ${\bf h}||
\hat{{\bf z}}$ and different values of exchange field.}
\label{fig:hz}
\end{figure}

Before illustrating our results for a F$\mid$NCS structure we
briefly comment on the proximity effect in an N$\mid$NCS structure since this has not been
studied previously in the literature. The qualitative features of the
DOS evaluated for this system strongly depend on the relative
amplitude of the singlet and triplet component: a qualitative change occurs when $\Delta_t$
exceeds $\Delta_0$ in magnitude. We find that the fully suppressed
low-energy DOS transforms into a peak-structure due to the
appearance of zero-energy states resulting from the triplet
component \cite{yokoyama_prb_05, iniotakis_prb_07}. In this way,
probing the DOS in a normal metal contacted to a NCS via \eg
STM-spectroscopy reveals direct information about the triplet
contribution in the superconducting condensate. We concentrate
here on the latter situation and fix the relative amplitude of gap
components as $\Delta_t=1.5\Delta_0$.


We now focus on the main result in this Letter: a spin-sensitive
long-ranged proximity effect. To understand how this effect works,
we first demonstrate it on two experimentally relevant examples: a
noncentrosymmetric superconductor described by
$\underline{\Delta}_\vk = \Delta_{0} \i\underline{\sigma_y} +
\mathbf{d}_\vk\cdot\boldsymbol{\underline{\sigma}}\i\underline{\sigma_y},\;
\mathbf{d}_\vk = \Delta_{t} \left(-k_y,k_x,0 \right)/k_F,$, and a
chiral $p_x+ip_y$-wave superconductor described by $\mathbf{d}_\vk
= \Delta_0(0,0,k_x+\i k_y)/k_F$. We choose these two particular cases because most experimental studies so far
suggest that they describe the superconducting states of CePt$_3$Si and
Sr$_2$RuO$_4$, respectively. In Fig. \ref{fig:hz}, the DOS
evaluated at the top of the structure $(x=d_F)$ and the exchange
field ${\bf h}|| \hat{{\bf z}}$ in the ferromagnet are shown. In
this situation, the zero energy peak (ZEP) in the DOS is
completely insensitive to the ferromagnetism in the NCS case (left
panel): even very for high exchange fields, the DOS remains
virtually unchanged compared to a normal diffusive electrode. The
peak is seen to persist even at distances far inside the
ferromagnetic layer $(\gg\xi_F)$, evidencing that we are dealing
with a long-range proximity effect in spite of the presence of an
exchange field. Moreover, we note that the proximity effect is
long-ranged \textit{even without any magnetic inhomogeneity}, in
complete contrast to conventional S$\mid$F hybrid structures. In
the case of a superconducting electrode featuring a chiral
$p$-wave symmetry (right panel), the situation changes
qualitatively: the proximity effect is now strongly dependent on
the exchange field, and vanishes completely when $h\gg\Delta_0$,
which is opposite to the NCS case.

\begin{figure}[t!] \centering \resizebox{0.46\textwidth}{!}{
\includegraphics{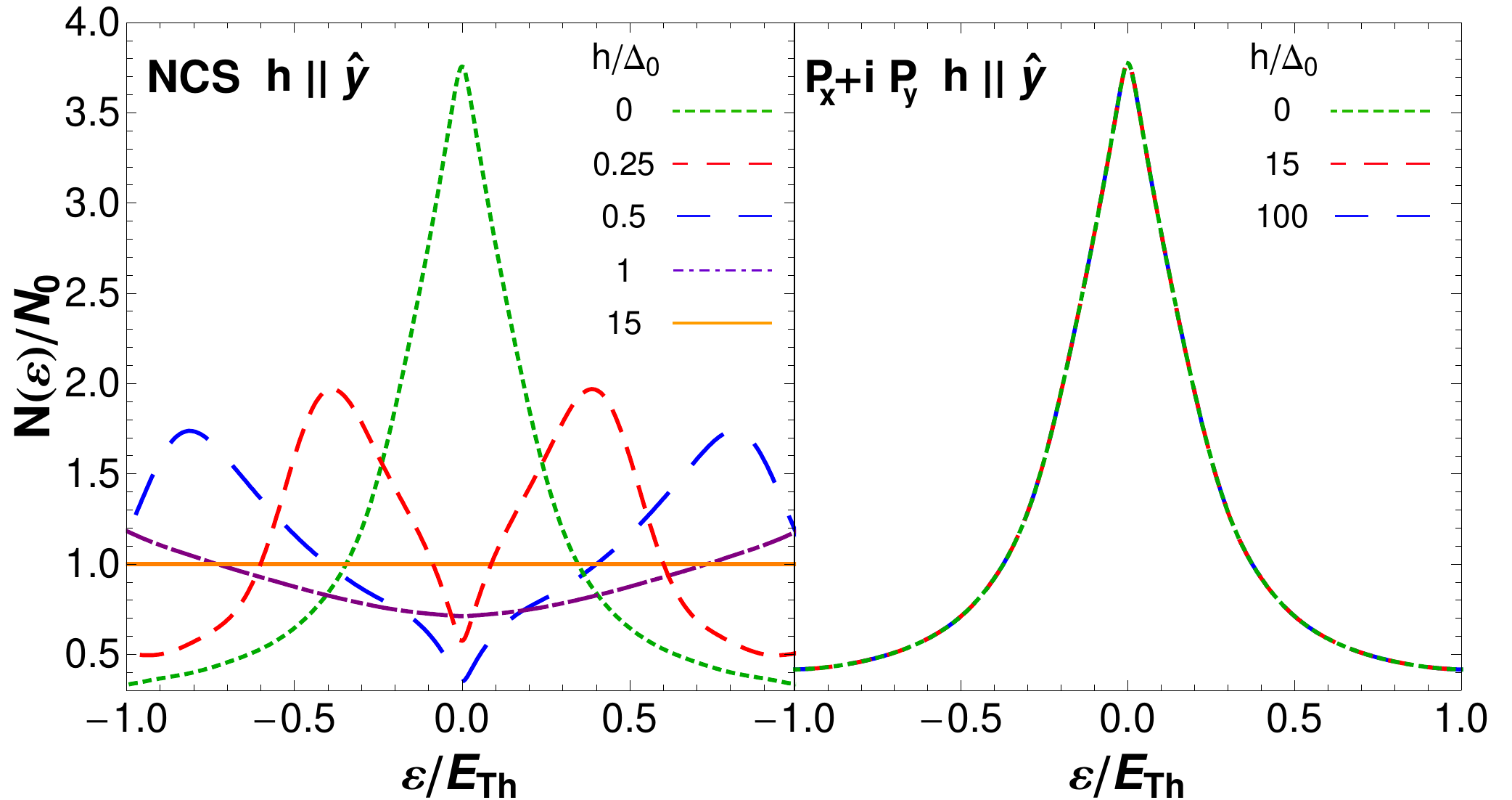}}
\caption{(Color online). {\it Left panel}: DOS at $x=d_F$ in a
F$\mid$NCS junction for ${\bf h}|| \hat{{\bf y}}$ and different
values of exchange field. {\it Right panel}: DOS at $x=d_F$ for
chiral $p_x+ip_y$ superconducting electrode for ${\bf h}||
\hat{{\bf y}}$ and different values of exchange field.}
\label{fig:hy}
\end{figure}

To demonstrate the spin-sensitivity of this proximity effect, we now turn to the case where the field satisfies
${\bf h}|| \hat{{\bf y}}$, as shown in Fig. \ref{fig:hy}. The field still lies in the plane of the ferromagnet, as is experimentally the case for \eg thin-film structures. In order to change the orientation from ${\bf h}|| \hat{{\bf z}}$ to ${\bf h}|| \hat{{\bf y}}$, one could grow two separate samples where the exchange field is locked to different orientations via \eg antiferromagnetic coupling, or alternatively grow the layers in different crystallographic orientations to effectively change the orientation of the exchange field.
As seen in Fig. \ref{fig:hy}, the
features of the DOS are now opposite to the case ${\bf h}|| \hat{{\bf z}}$: the
peak appearing for the NCS is destroyed by the exchange field while the peak in the
chiral $p_x+\i p_y$ case remains uninfluenced by the presence of an exchange field. As before, this not only leads to the survival of the ZEP in the DOS, but also makes the proximity effect long-ranged inside the ferromagnet even if there is no magnetic inhomogeneity. In effect, the roles of the NCS and the chiral superconductor have been reversed. These results suggest that simply by altering the in-plane orientation of the exchange field in the ferromagnet, the proximity-induced DOS serves as a clear discriminator against different triplet states.

We now proceed to explain the underlying physics behind these results. The triplet pairing in the superconducting condensate has a spin-degree of freedom described by the $\mathbf{d}_\vk$-vector and thus couples to the orientation of the exchange field. In the NCS case, the net triplet condensate has zero spin-projection, but it is comprised of two equal contributions of $S_z=\pm1$ pairing gaps. Therefore, an exchange field applied along the $z$-direction has no influence on the triplet pairing since it simply renormalizes the chemical potential. However, applying the field along the $y$-direction effectively induces spin-flip scattering which breaks the $S_z=\pm1$ Cooper pairs and thus suppresses the proximity effect. Because of this, the ZEP is present for $\mathbf{h}||\hat{{\bf z}}$ but absent for $\mathbf{h}||\hat{{\bf y}}$. In the chiral superconducting state, chosen to model the pairing symmetry believed to be realized in Sr$_2$RuO$_4$, the triplet component belongs to the $S_z=0$ class and thus has a spin confined to the $xy$-plane. By a similar argument as above, an exchange field applied in this plane does not interact destructively with the triplet pairing, while a field applied along $\hat{{\bf z}}$ induces spin-flip scattering detrimental to the Cooper pairs.

This line of reasoning holds quite generally for any triplet state. Although there exist alternative experimental techniques to probe the triplet pairing symmetry which are also based on the interaction between a magnetic field and the $\mathbf{d}_\vk$-vector, such as spin susceptibility and thermal conductance measurements \cite{maeno_review, matsuda_review}, these techniques are based on applying an external magnetic field directly to the superconductor. In contrast, the spin-sensitive proximity effect considered in this Letter has an advantage compared to previous methods in that the superconducting and magnetic correlations originate with different parts of the system and that the measurements do not have to be performed on the superconductor, thus avoiding complications with induced vortex-configurations via an external field. Instead, all the information about the triplet condensate can be probed simply by measuring the DOS in the non-superconducting region. The proximity effect also becomes long-ranged in the ferromagnet irrespective of whether it has a weak or strong exchange field, such that one is not restricted to probing the correlations within a few nanometers of the interface region as in conventional S$\mid$F structures.

Concerning the experimental realization of the
ferromagnet$\mid$triplet superconductor junctions considered here,
one would need to couple the ferromagnetic layer to the plane of
the superconductor where the order parameter experiences a sign
reversal, in order to produce the ZEP. Depending on the crystal
structure, this could be challenging with respect to cleaving
along an appropriate lattice plane for certain materials, \eg the
$ac$-plane for Sr$_2$RuO$_4$. However, we also note that a
possible pinning of the $\mathbf{d}_\vk$ vector by
spin-orbit coupling in triplet superconductors \cite{maeno_review}
is not only unproblematic for our purposes, but actually
beneficial since it stabilizes the orientation of the
$\mathbf{d}_\vk$ vector. We have also checked numerically how
robust the spin-sensitive proximity effect is against a
misalignment of the field, i.e. when it is not fully oriented
along the $y$- or $z$-axis for our setup, due to \eg the presence
of stray fields or magnetic inhomogeneities. Only small
quantitative changes are observed for misalignment angles up to
15-20$^\circ$, such that the effect exhibits some robustness
towards imperfections in the orientation of the field. Finally, we note that our results could be used to identify if the $\mathbf{d}_\vk$-vector in 
eutectic Sr$_2$RuO$_4$-Sr$_3$Ru$_2$O$_7$ is stabilized to $(-k_y,k_x,0)$ instead of
$(0,0,k_x+\i k_y)$ as in bulk Sr$_2$RuO$_4$, which was suggested very recently in Ref. \cite{yanase}.

In summary, we have demonstrated the presence of a spin-sensitive proximity effect in ferromagnet$\mid$triplet superconductor bilayers. In particular, we have shown how the orientation of the exchange field in the ferromagnet couples to the $\mathbf{d}_\vk$-vector of the proximity-induced triplet superconductivity leaking into the ferromagnet region, providing clearly distinguishable features in the local density of states. With the ongoing activity of characterizing novel superconducting materials where triplet pairing is believed to be present, such as heavy-fermion compounds, we believe our results may serve as a useful tool to experimentally identify the superconducting pairing state.

\textit{Acknowledgments}. We thank Y. Tanaka, Y. Asano, and T. Yokoyama for
helpful comments.

\end{document}